\documentstyle[manuscript,aps]{revtex}
\begin{document}
\draft
\title {Possible doublet mechanism for a regular component of parity
violation in neutron scattering}
\author{{\bf V.V.Flambaum$^{1}$ and V.G.Zelevinsky$^{2}$}}

\address{$^{1}$School of Physics, University of New South Wales\\
Kensington, New South Wales 2033, Australia\\
$^{2}$National Superconducting Cyclotron Laboratory\\
 and\\
 Department of Physics and Astronomy \\
Michigan State University, East Lansing, MI 48824, USA\\}
\maketitle

\begin{abstract}
A nucleus with octupole deformation of the mean field reveals
rotational doublets with the same angular momentum and opposite parity.
Mediated by the Coriolis-type interaction, the doublet structure
leads to a strong regular component in the parity violation caused by
weak interaction. This can explain sign correlations observed in
polarized neutron scattering by $^{232}$Th.
\end{abstract}


Large effects of parity nonconservation (PNC) in compound nuclear states
\cite{Alf,Mas,Bow} are observed in experiments with polarized neutrons. Now
it is widely accepted that the mechanism of statistical
(sometimes called "dynamical") enhancement
of weak interactions due to the high level density in compound
nuclei \cite{Haas,BS,Shap,SF,Kad,AAB} is responsible for the large magnitude of
the effects.
Roughly speaking, the enhancement in comparison to the parity
mixing between single-particle states is proportional to $\sqrt{N}$
where $N \sim 10^{5}-10^{6}$ is a number of simple shell
model configurations in a generic complicated compound wave function.
Matrix elements of the weak interaction $H_{W}$ between such states
of opposite parity are suppressed by a factor $\sim 1/\sqrt{N}$ with
respect to single-particle estimates but the level spacing in the
denominator is diminished by factor $\sim N$. It results in statistical
enhancement $\sim \sqrt{N} \sim 10^{3}$.

Additional enhancement due to the interference of low-energy
$s$-wave and $p$- wave neutron resonances mixed by weak interaction
is provided\cite{SF} by the factor $(\Gamma_{s}^{(n)}
/\Gamma_{p}^{(n)})^{1/2} \sim 1/kR \sim 10^{2}-10^{3}$
where $k$ is neutron wave number, $R$ is nuclear radius,
$\Gamma_{s}^{(n)}$ and $\Gamma_{p}^{(n)}$ are neutron widths
for the $s$- and $p$-waves, respectively. Acting together, those factors
of statistical and kinematical enhancement
lead to the longitudinal asymmetry (or analyzing power measuring
the difference of total cross sections $\sigma_{\pm}$
for neutrons with positive and negative helicity)
\begin{equation}
P = \frac{\sigma_{+}-\sigma_{-}}{\sigma_{+}+\sigma_{-}}   \label{1}
\end{equation}
which reaches in some cases the magnitude of about 10\%.

Statistical nature of the enhancement implies
randomness of matrix elements of weak interaction between
complicated states of opposite parity. Therefore one would expect
the random sign of asymmetry (1). Recent LAMPF experiments
\cite{Bow} show, contrary to this expectation, that neutron capture
to $p$-wave compound resonances in $^{233}$Th leads to the asymmetry
of the same sign for all eight resonances where the effect is statistically
significant. At the same time, other compound nuclei seem to
be closer to the
random distribution of the sign  of asymmetry. In all cases the
order of magnitude of the effect agrees with the predictions based
on the abovementioned two factors. Therefore several attempts have
been made \cite{Fla,Bow1,Auer,Lew,Gud} to find a mechanism
which would be able to
generate a regular (non-random) component of the PNC asymmetry. These
attempts were unsuccessful in the sense that the desired magnitude
of the weak matrix element was inconsistent with our knowledge
on weak interactions in nuclei.

The experimental pattern for a target of $^{232}$Th
gives a hint that this nucleus, and consequently the compound nucleus
$^{233}$Th, might be a special case due to some pecularities
of its structure as compared to "normal" deformed heavy nuclei like
$^{238}$U which apparently exhibit random asymmetry. Below we discuss the
specific features of parity violation which could be related to
structure of Th nuclei.

It is known that Th isotopes display strong octupole correlations
\cite{BM,Leander,Ots,Naz,Jol}.
The potential energy surface in the space of the quadrupole
and octupole deformation parameters is quite complicated. As a function
of quadrupole deformation it has triple-humped structure \cite{Blons,Cw}.
Due to the mass asymmetry of fission, on the top of the outer barrier
the cold nucleus goes through the pear-shaped configurations
(fission channels \cite{BM}) corresponding to large octupole deformation.
We assume that static octupole deformation in Th is present already in the
first (ground state) potential well. However, this assumption is not critical
because at the excitation energies near neutron threshold, the nuclear
wave function in the space of the deformation parameters certainly has a
significant probability of large octupole deformation, $\beta_{3}\simeq 0.35$
\cite{Pash,Bengt,Cw}.

For sufficiently strong deformation, the adiabatic approximation is
justified which allows one to write down \cite{BM} the wave functions as
products of orientational $D$-functions and intrinsic functions $|\chi\rangle$.
In the case of axial symmetry the projection $K = {\bf In}$ of the
total angular momentum ${\bf I}$ on the intrinsic symmetry axis ${\bf n}$
is conserved and can be used to label the intrinsic wave
function $|\chi\rangle = |a;K\rangle$. In neutron capture by
a spinless target we are interested in states with $|K| = 1/2$.
For a given intrinsic function with $K\neq 0$ presence of octupole
deformation, or of any shape which is axially symmetric but has
no symmetry with respect to reflection in the equatorial plane,
leads \cite{BM} to rotational doublets with definite parity $\eta$,
\begin{equation}
|\Psi^{I}_{MK;\eta}\rangle = \sqrt{\frac{2I+1}{8\pi}}\bigl\{D^{I}_{MK}(\varphi,
\theta,0)|a;K\rangle + \eta (-1)^{I+K} D^{I}_{M-K}(\varphi,\theta,0)
|a;-K\rangle \bigr\}.                                     \label{2}
\end{equation}
Energy splitting of doublet states implies that there exist a physical
interaction which can couple "right- and left- oriented" configurations
$|a;\pm K\rangle$. One can imagine various specific mechanisms of
coupling\cite{SFY,SF}, for example tunneling of an excess cluster. In
the case of $K=1/2$ the Coriolis force acting in the first
order can be sufficient to generate this coupling similar to
the decoupling parameter in the normal spectra of odd-$A$ deformed nuclei
\cite{BM}.

Since the energy separation within the doublet is supposedly of the
order of several keV one can expect that mixing of the doublet states
of opposite parity by weak interaction is much more favorable than
mixing of the single-particle orbitals separated by MeV. However,
the direct mixing of states (2) with the same intrinsic structure
and opposite $\eta$ is possible only if the weak perturbation $H_{W}$ violates
time reversal symmetry along with space inversion symmetry \cite{SFY} (see
also \cite{BM,SF}). Indeed, the mixing matrix element can be expressed in
 terms of intrinsic expectation values of weak interaction,
\begin{equation}
\langle \Psi^{I}_{MK;-\eta}|H_{W}|\Psi^{I}_{MK;\eta}\rangle = \frac{1}{2}
\{\langle a;K|H_{W}|a;K\rangle - \langle a;-K|H_{W}|a;-K\rangle \}.
                                                          \label{3}
\end{equation}
Since $H_{W}$ is a pseudoscalar, the intrinsic matrix element
$\langle a;K|H_{W}|a;K\rangle$ should
be proportional to the intrinsic pseudoscalar $K$. On the other hand,
it means that this quantity, together with $K$, changes sign under
time reversal which contradicts to $T$-invariance of $H_{W}$. Actually,
one can see from (3) that only $T$-odd interaction leading to the
opposite sign of the two matrix elements in curly brackets can
result in the direct mixing of the doublet states.

Note that for a $P$- {\sl and} $T$-odd interaction direct mixing within
the doublet is not forbidden which can be of some interest for the
problem of search for $T-$ and $P$-violating nuclear forces. In this
respect the situation is similar to that in the problem of the
electric dipole moment of a particle \cite{Lan}.

To clarify the situation we can refer to the simple model where the
mixing of single-particle $s$- and $p$-orbitals is caused by an electric
field through the dipole interaction $-{\cal E}z$ as for an electron
in a dipole molecule. The mixed orbitals $|\pm 1/2 \rangle$ with the
projection $j_{z} =m = \pm (1/2)$ of the particle angular momentum onto
the field axis are
\begin{equation}
|\pm 1/2\rangle = \sqrt{1-\xi^{2}_{\pm}}|s_{1/2},\pm\rangle +
\xi_{\pm}|p_{1/2},\pm\rangle                                      \label{4}
\end{equation}
where the polarizability coefficients $\xi_{\pm}$ differ merely by sign,
$\xi_{\pm} = \pm \xi$. Indeed, the spin-angular structure of
the $p_{1/2}$ wave function is $\sim (\vec{\sigma}{\bf r})
|s_{1/2}\rangle$. Therefore the polarizability is proportional to the diagonal
matrix element $\langle s_{1/2}|z(\vec{\sigma}{\bf r})|s_{1/2}\rangle$
which, in turn, is proportional to the matrix element of $j_{z}$ and
changes its sign for $m \rightarrow - m$. Now, the expectation value
\begin{equation}
\langle \pm 1/2|H_{W}|\pm 1/2 \rangle = \pm 2\xi \sqrt{1-\xi^{2}}
{\rm Re} \langle s_{1/2},\pm|H_{W}|p_{1/2},\pm\rangle                 \label{5}
\end{equation}
vanishes for the standard parity-nonconserving weak interaction when
the amplitude $\langle s|H_{W}|p \rangle$ is an imaginary pseudoscalar.
Contrary to that, the similar amplitude for a hypothetical $P$- and
$T$-violating hamiltonian $H_{PT}$ would be a real pseudoscalar so that
eq.(5) would give non-zero expectation values of the same magnitude
but of the opposite sign for the projections $\pm 1/2$.

Going back to our original problem we see that
the mixing of the doublet states by the $P$-odd but
$T$-even interaction should be mediated by another ("normal", $P$-
and $T$-even) interaction $H'$ leading to the non-adiabatic admixtures
of different configurations $|b;K'\rangle$.
In particular, it can (but does not have to)
be the same interaction which was already mentioned as inducing the
energy splitting {\sl within the doublet}. The interaction $H'$
in the first order
influences PNC via matrix elements $\langle a;-K|H'|b;K\rangle$ which
appear in the $P$-conserving mixing matrix element
\begin{equation}
\langle \Psi_{MK;\eta}^{aI}|H'|\Psi_{MK;\eta}^{bI}\rangle =
\eta A_{IK}\langle a;-K|H'|b;K\rangle                    \label{6}
\end{equation}
with the amplitude $A_{IK}$ depending on the nature of interaction $H'$.
As a result of the joint action of $H'$ and $H_{W}$ the total rotational
function (1) acquires an admixture of opposite parity,
\begin{equation}
|\Psi^{aI}_{MK;\eta}\rangle \rightarrow |\tilde{\Psi}^{aI}_{MK;\eta}\rangle
= |\Psi^{aI}_{MK;\eta}\rangle + \beta |\Psi^{aI}_{MK;-\eta}\rangle,
                                                           \label{7}
\end{equation}
where the mixing amplitude $\beta$ is
\begin{equation}
\beta = -2\eta \frac{A_{IK}}{E-E_{-\eta}}\sum_{b}\frac{\langle a;-K|
H'|b;K\rangle \langle b;K|H_{W}|a;K\rangle}{E - E_{b}}.   \label{8}
\end{equation}
Here $E$ is neutron energy and, up to our accuracy,
in the denominator of the sum we neglect
the rotational splitting of the doublet $b$.
This admixture is directly related to the observed PNC asymmetry (1),
\begin{equation}
P = 2 \sqrt{\frac{\Gamma^{(n)}_{s}}{\Gamma^{(n)}_{p}}}\beta.     \label{9}
\end{equation}
If the splitting is
due to the same interaction $H'$ it also can be expressed in terms
of the amplitude $A_{IK}$,
\begin{equation}
E_{\eta}^{aI} - E_{-\eta}^{aI} = 2\eta A_{IK}\langle a;-K|H'|a;K \rangle.
                                                             \label{10}
\end{equation}
In this case the resulting PNC admixture (8) at the resonance
energy $E\approx E_{\eta}$ does not depend on $A_{IK}$.

Note that in the sum (8) the numerator contains two matrix elements
and both of them are suppressed $\sim 1/\sqrt{N}$ for generic compound
wave functions $a$ and $b$. Therefore the contribution of the closest
states with the energy difference of the order of the mean level spacing $D$ in
the compound nucleus is not statistically enhanced. Then one has take
into account contributions of distant states $b$. If the product
of matrix elements is peaked for the states $b$ on the distance $E_{b}-E
\simeq \omega$ from the resonance, the closure approximation gives
\begin{equation}
\beta = 2\eta \frac{A_{IK}\langle a;-K|H'H_{W}|a;K\rangle}{\omega
(E - E_{-\eta})},                                        \label{11}
\end{equation}
and, in the case (10), at $E\approx E_{\eta}$ we come to a remarkably
simple result
\begin{equation}
\beta = \frac{\langle a;-K|H'H_{W}|a;K\rangle}{\omega \langle a:-K|
H'|a;K\rangle}.                                         \label{12}
\end{equation}
Thus, in this scheme one can expect the admixture amplitude of the order
$\beta \simeq (H_{W})_{s-p}/\omega$ where $(H_{W})_{s-p}$ is given by
the typical single-particle matrix elements of weak interaction,
$(H_{W})_{s-p} \simeq 5$ eV. For the Coriolis force as $H'$, the
transition energy between deformed single-particle orbitals
with $\Delta m = \pm 1$ is of the order of 100 keV which leads to
the estimate $\beta \simeq 5\times 10^{-5}$.

This can be compared with
the mixing between compound states of opposite parity which can
be estimated as the ratio of the typical corresponding matrix element to the
level spacing, $\beta_{comp}\simeq \langle H_{W}\rangle_{comp} /D$.
The experimental data \cite{Bow} show that $\langle H_{W}\rangle_{comp}
\approx 1.3\times 10^{-3}$ eV. Mixing matrix elements of the same order
are predicted by theoretical calculations \cite{Vor} using the gas of
thermally excited quasiparticles as a model for a compound nucleus. Such
an approach gained a support in the recent detailed analysis
\cite{temp} of chaotic
shell model wave functions. Using this value and $D=15$ eV, we obtain
$\beta_{comp} \simeq 10^{-4}$.

As a result, the
estimate of the regular component of the observable effect (9) gives
$P\simeq 10^{-2}$. It means that the regular effect due to the doublet
mechanism has to be seriously taken into consideration.

Our arguments are based on the assumption that the pear shape
and related doublet structure persist at required
excitation energies. If this is the case, the complicated intrinsic states
are the superpositions
\begin{equation}
|a;\pm K\rangle = \sum_{i} C_{i}^{a} |\Phi_{i};\pm K\rangle
                                                          \label{13}
\end{equation}
of simple quasiparticle configurations $|\Phi_{i}; \pm K\rangle$
with amplitudes $C_{i}^{a}$ independent of the sign of $K$.
Therefore the matrix element in (13) contains a regular contribution
\begin{equation}
\langle a;-K|H'H_{W}|a;K\rangle \approx \sum_{i}(C^{a}_{i})^{2}
\langle \Phi_{i};-K|H'H_{W}|\Phi_{i};K\rangle.      \label{14}
\end{equation}
As mentioned above, such expressions (see also energy difference (10))
 can be calculated explicitly if
the excited nucleus is modeled by a gas of quasiparticles which, in
the case under study, are moving in the pear-shaped field. Sums with
the weights $C^{a}_{\alpha}$ are substituted in this approach by the
expectation values for the thermal equilibrium ensemble \cite{Vor,temp}.

Moreover, this approach is applicable regardless of the presence
or lack of axial symmetry. In the general case the wave functions (2) are still
similar combinations of the "right" and "left" states and we
can view $\pm K$ in (13) as a general label necesssary to distinguish
mutually reflected wave functions.

Different consideration of the regular
 contribution of the doublet states  was independently started in \cite{ABS}.

The authors acknowledge hospitality and support from the INT, Seattle,
where part of this work was done. V.Z. would like to acknowledge support
from the NSF grant 94-03666. V.F. is grateful to G.I. Mitchell who
showed preliminary results of Los Alamos PNC measurements prior to
publication and initiated this work. He also would like to
acknowledge support from ARC and DITAC.

\end{document}